\documentclass[aps,prl,twocolumn,10pt,longbibliography]{revtex4-1} 

\usepackage{graphicx}  
\usepackage{amsmath}
\usepackage{times}
\usepackage[T1]{fontenc}
\usepackage{mdframed}
\usepackage{epstopdf}
\usepackage{soul}
\usepackage[T1]{fontenc}
\usepackage[utf8]{inputenc}
\usepackage[english]{babel}
\usepackage{blindtext}

\begin{document}

\title{Experimental Observation of Large Chern numbers in Photonic Crystals}
\author{Scott A.~Skirlo$^1$}
\email{sskirlo@mit.edu}
\author{Ling ~Lu$^1$}
\email{linglu@mit.edu}
\author{Yuichi Igarashi$^{1,2}$}
\author{Qinghui Yan$^{1,3}$}
\author{John Joannopoulos$^1$}
\author{Marin ~Solja\v{c}i\'{c}$^1$ \\ $^1$ Department of Physics, Massachusetts Institute of Technology, Cambridge, MA 02139, USA\\
                                                        $^2$ Smart Energy Research Laboratories, NEC Corporation, 34 Miyukigaoka, Tsukuba, Ibaraki 305-8501, Japan \\
                                                         $^3$ The Electromagnetics Academy at Zhejiang University, Zhejiang University, Hangzhou 310027, China \\
                                                        }

\date{\today}

\begin{abstract}


Despite great interest in the quantum anomalous Hall phase and its analogs, all experimental studies in electronic and bosonic systems have been limited to a Chern number of one. Here, we perform microwave transmission measurements in the bulk and at the edge of ferrimagnetic photonic crystals. Bandgaps with large Chern numbers of 2, 3, and 4 are present in the experimental results which show excellent agreement with theory. We measure the mode profiles and Fourier transform them to produce dispersion relations of the edge modes, whose number and direction match our Chern number calculations.



\end{abstract}

\maketitle


The Chern number~\cite{tknn} is an integer defining the topological phase in the quantum Hall effect (QHE)~\cite{klitzing1980new}, which determines the number of topologically-protected chiral edge modes. The quantum anomalous Hall effect (QAHE) possesses these same properties as an intrinsic property of the bandstructure with time reversal symmetry breaking~\cite{haldane1988model,lu2014topological}. Recent experiments have discovered the QAHE and its analogues in ferrimagnetic photonic crystals~\cite{wang2009observation}, magnetically-doped thin films~\cite{chang2013experimental} and in ultracold fermion lattices~\cite{jotzu2014experimental}. However, the Chern numbers observed in all of these systems were limited to $\pm{1}$. Finding larger Chern numbers would fundamentally expand the known topological phases~\cite{jiang2012quantum,wang2013quantum,fang2014large,mytheorypaper,perez2015}. 

Here, we provide the first experimental observation of Chern numbers of magnitude 2, 3 and 4, by measuring bulk transmission, edge transmission, and the edge mode dispersion relations in a ferrimagnetic photonic crystal.  The excellent agreement between the experiment and modeling allows us to identify various topological bandgaps and map out the dispersion relations of one-way edge modes for the first time in \textit{any} QHE or QAHE system in nature.


In a 2D system one can realize bands with nonzero Chern numbers, and generate the QAHE, by applying a T-breaking perturbation \cite{haldane2008possible,raghu2008analogs,wang2008reflection}. The Chern number is defined as the integral of the Berry flux over the entire Brillioun zone. When connected bands are gapped by a T-breaking perturbation, the bands will exchange equal and opposite Berry flux at each degenerate point, with the total Berry flux exchanged determining the Chern number. For instance, two isolated bands connected by one pair of Dirac points gapped by T-breaking will acquire $\pm2\pi$ Berry flux ($\pi$ from each Dirac point) and a Chern number associated with the bandgap (``gap Chern number'') of $\pm 1$. A general way to calculate the gap Chern number ($C_{gap}=\Sigma C_{i}$) is to sum the Chern numbers of all the bands below the bandgap~\cite{Fukui}. A bandgap with $C_{gap}=0$ is trivial, while a bandgap with $C_{gap}\neq{}0$ is topologically nontrivial. 

In our previous theoretical study we found that the magnitude of the gap Chern number can be increased above one by simultaneously gapping multiple sets of Dirac and quadratic degeneracies. If Berry flux from the gapped degeneracies adds constructively, $C_{gap}$ can be large. In Fig. 1a we present a theoretical topological gap map for a 2D ferrimagnetic photonic crystal as a function of the externally applied magnetic field and the frequency, showing nontrivial bandgaps with $C_{gap}$ from $-4$ to $3$. We studied this same square lattice in an experiment, to verify these predictions. 


\begin{figure*}
\centering
		\includegraphics[width=\textwidth]{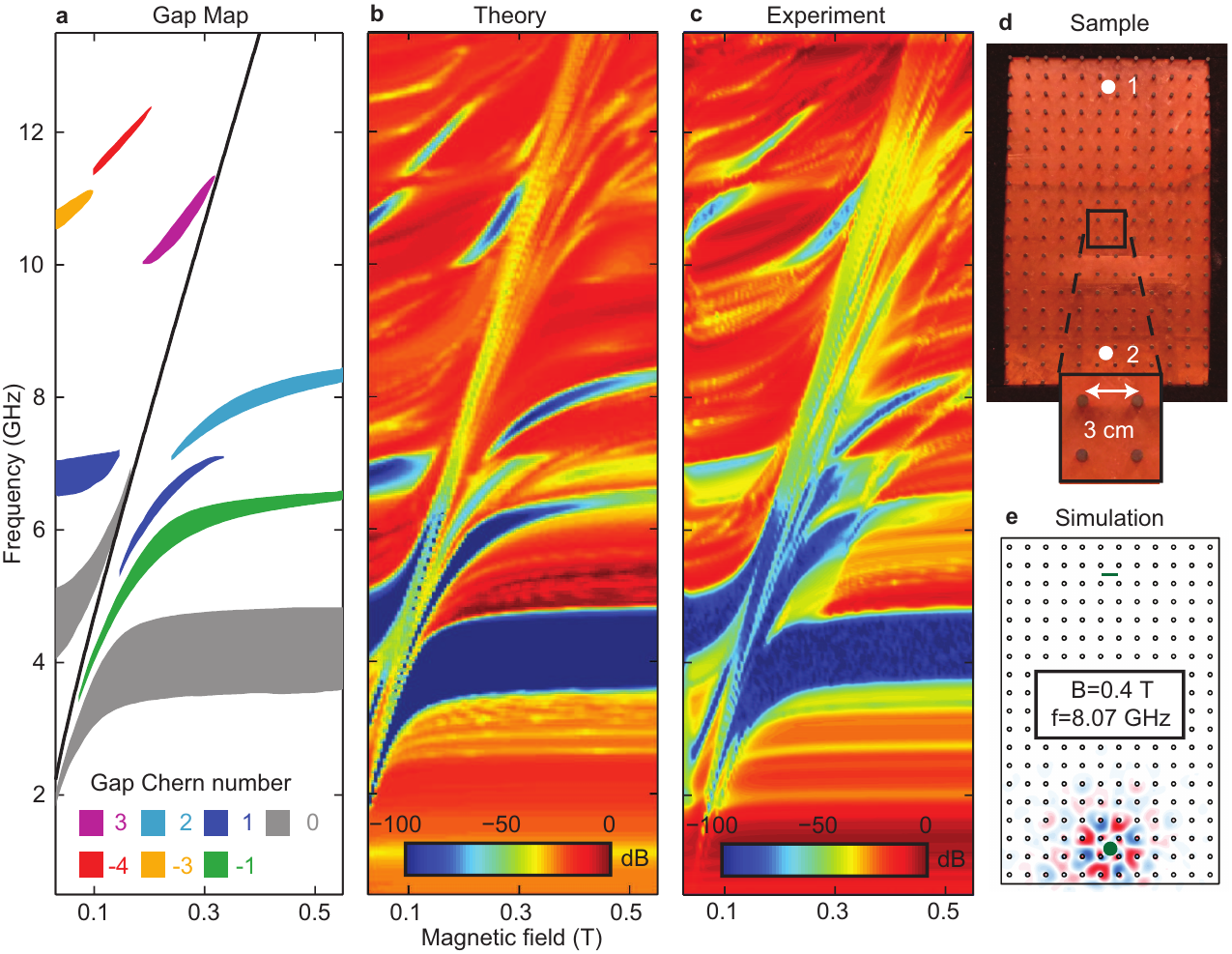} 
	\caption{Comparison of theoretical gap map and bulk transmission to experimental transmission measurement in a 2D ferrimagnetic photonic crystal. a) Theoretical topological gap map as a function of the magnetic field and the frequency with each bandgap labeled by its gap Chern number. The diagonal black line indicates the resonance in the effective permeability (Supplementary Material) b) Theoretical bulk transmission c) Experimental bulk transmission d) Experimental configuration with the lattice geometry (top metal plate removed). The antenna locations are marked with ``1'' and ``2''. e) Simulation geometry with the green line representing the receiving antenna, and the green circle representing the transmitting antenna. }
	\label{fig:two-edge-modes}
\end{figure*}

The experimental configuration resembles a prior work which demonstrated $|C_{gap}|=1$~\cite{wang2009observation}. A square lattice of ferrimagnetic garnet rods is placed between two conductive copper plates. This configuration forms a parallel-plate waveguide, with the electric field perpendicular to the plate. Since the electric field for the fundamental mode is constant as a function of height, this is equivalent to a 2D system.  The modes in the photonic crystal are excited by antennas attached to the top plate and fed to a network analyzer. Around the boundary of the system we placed an absorber to minimize reflections and outside interference. We include an overhead image of one of the crystals we constructed in Fig. 1d. 

To observe the QAHE analog in the experiment, we break T-symmetry by applying a spatially uniform magnetic field to the ferrimagnetic garnet rods, which acquire off-diagonal imaginary parts in the permeability tensor~\cite{pozar1998microwave}. Unlike electrons, the external magnetic field does not interact directly with photons. However, for this system, Maxwell's equations can be written in the exact same form as the Schrodinger equation with a periodic vector potential~\cite{wang2008reflection}. This makes the system an analogue of the QAHE. Our photonic crystals were placed in the MIT cyclotron magnet, and the magnetic field was swept between 0.03 T and 0.55 T to characterize the transmission of the photonic crystal as a function of the magnetic field and the frequency. 

We show the experimental transmission through a bulk photonic crystal in Fig. 1c. Here the color illustrates the amplitude of the transmission between the antennas in decibels ($S12=20\log\frac{E_{in}}{E_{out}}$). In the plots there are several deep blue regions of low transmission that clearly correspond to the locations of bandgaps in the gap map. The sweeping feature that extends diagonally across the figure is due to the gyromagnetic resonance of the ferrimagnetic garnet rods (Supplementary Material). The resonant frequency of the effective permeability is plotted with a black line in Fig. 1a.   

In Fig. 1b we present the corresponding theoretical data for transmission through a lattice of the same size and dimension calculated with COMSOL. One of the transmission simulations is shown in Fig. 1e. For direct comparison, the transmission data in Fig. 1b is plotted with the same colorbar scale as the experiment in Fig. 1c. The slight offset of about 0.04 T in the magnetic field between the theoretical and experimental plots is caused by demagnetization (Supplementary Material). Clearly the theoretical and experimental transmission bear strong resemblance to each other and the topological gap map, showing that a square lattice of ferrimagnetic rods can contain a wide variety of different $C_{gap}$ numbers. 

\begin{figure*}
\centering
		\includegraphics[width=\textwidth]{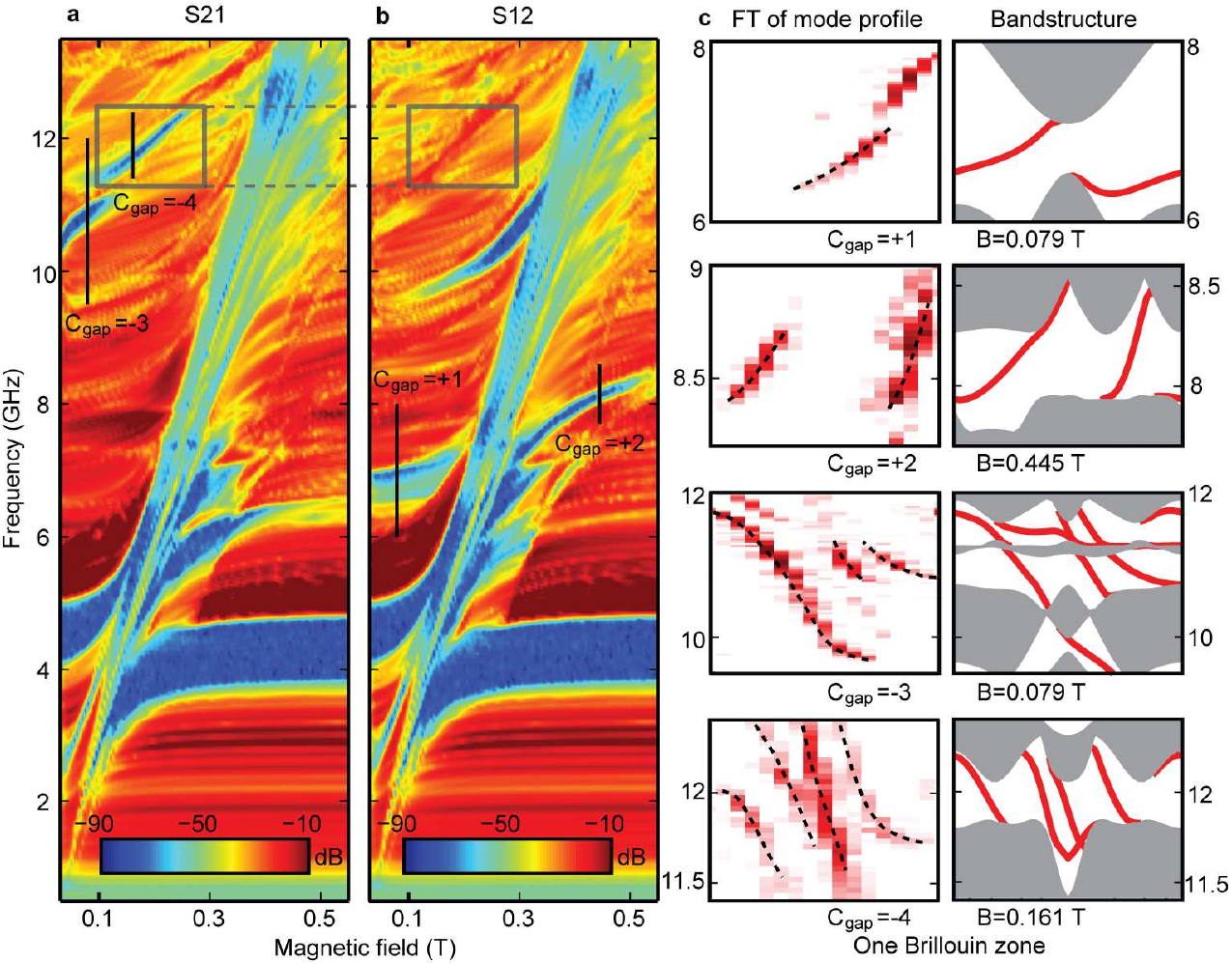} 
	\caption{Experimental edge transmission measurement and Fourier transform (FT) of mode profiles along the copper boundary. a) S21. b) S12. The bandgaps that are nontrivial have direction-dependent edge transmission, because the interface of a nontrivial bandgap with a trivial bandgap (copper boundary) supports one-way modes. In a) and b) this causes the nontrivial bulk bandgaps from Fig. 1c to be present in one direction (e.g. S12) and be absent in the other (e.g. S21), which we highlight for the $C_{gap}=-4$ bandgap with black boxes. The trivial bandgaps around 4 GHz do not support one-way modes, and so do not exhibit direction-dependent transmission. c) Experimental FT of edge mode profiles and the theoretical edge bandstructures with the edge modes in red and the bulk bands in gray. The range of wavevectors included in both plots is the same and includes only one Brillouin zone. The number of one-way edge modes in both sets of plots agrees with $|C_{gap}|$ from Fig. 1a, while the sign of $C_{gap}$ is consistent with the theoretical group velocity (from the edge mode dispersion) and the directional transmission in a) and b). }
	\label{fig:two-edge-modes}
\end{figure*}

Several nontrivial bandgaps in Fig. 1 and in the supplementary material occur even at low magnetic fields. This indicates that topological effects can be achieved at low applied magnetic fields($\sim{}$0.03 T) enabling various studies and applications. Furthermore, these same bandgaps would remain open at zero external magnetic field by using ferrimagnetic materials with remanent magnetization~\cite{green1974microwave}; this way, a future experiment could be performed even without external magnetic fields.

One-way edge modes are present at the boundary between two crystals with a nontrivial and a trivial bandgap respectively, or at the boundary between crystals with nontrivial bandgaps with different $C_{gap}$~\cite{hatsugai1993chern,lu2014topological}. If the bandgaps of two neighboring crystals overlap in frequency, the number of edge states in the shared frequency gap is determined by the difference between the gap Chern numbers of each crystal. The sign of this difference determines the directions of the edge states. This means that with the nontrivial bandgaps we found, constructing one-way waveguides with up to seven modes is possible. If one of the materials is trivial ($C_{gap}=0$), like metal or air, the number of edge states equals the gap Chern number of the crystal, with the sign of this number determining their directions. 

 To provide more evidence of the topological state of these bandgaps and the one-way modes we modified the setup to include a highly conductive copper boundary at the edge of the crystal. This boundary acts as a mirror with a trivial bandgap. We place two antennas near this edge on each side of the sample and measure the transmission between them. In Fig. 2a and Fig. 2b we present both the S12 and S21 parameters to describe the direction-dependent transmission of the edge modes along the metal boundary. S12 refers to exciting the second antenna and measuring with the first antenna, while S21 is the opposite. 

The bandgaps that are nontrivial ($C_{gap}\neq{}0$) can be identified in Fig. 2 because they will have direction-dependent edge transmission. Specifically the nontrivial bandgaps measured in Fig. 1c will appear in either Fig. 2a or b, but not both. We show this explicitly for the $C_{gap}=-4$ bandgap by highlighting the direction-dependent transmission with gray boxes. This arises from the directional edge states as follows. In one direction, the group velocity of the edge modes is opposite that required to travel to the receiving antenna, so the transmission measurement will record the bulk bandgap. However, in the other direction, the group velocity of the edge modes is in the same direction as is required to get to the receiving antenna, so the bandgap will appear to be nonexistant. Trivial bandgaps ($C_{gap}=0$) around 4 GHz do not support one-way edge modes, and so do not exhibit direction-dependent transmission at the edge. 

To further study the topological nature of these bandgaps we measured the mode profile at the edge of the photonic crystal. We accomplished this by mounting one antenna for excitation to the lower plate, and another small dipole antenna for measurement to the upper plate~\cite{li2014}. During the measurement, the upper plate was translated a total of 47 cm in 2.5mm steps. At each step both the phase and amplitude of the electric field was recorded (Supplementary Material). From this spatial data the mode profile in the waveguide can be reconstructed. The Fourier transform of the mode profile produces the dispersion relation of the waveguide which we present on the left-hand side of Fig. 2c. 

In Fig. 2c on the right-hand side we include a comparison with the edge band calculations with the bulk bands in gray, and the edge modes in red. 
It is clear that the calculated edge-mode dispersion shows an excellent agreement with the dispersion relations extracted from experiments. The number of edge modes is equal to the gap Chern number for each inset. The sign of $C_{gap}$ is consistent with the group velocity of the edge modes and agrees with the directional edge transmission data in Fig. 2a and 2b. In the supplementary material we present additional simulations validating these results for $C_{gap}=-3$ and $-4$. These results consistute the first direct measurement of one-way edge mode dispersion in any QHE system. 



To further study the gap Chern numbers of the observed topological bandgaps, we construct a topological one-way circuit~\cite{mytheorypaper}. As illustrated in Fig. 3d, this consists of a $C_{gap}=2$ (a=3.0 cm) crystal and $C_{gap}=1$ (a=2.4 cm) crystal, with a copper boundary on the edge. We present the design and calculations characterizing the $C_{gap}=1$ crystal in the Supplementary material, while the results from the $C_{gap}=2$ crystal are shown in Fig. 1 and Fig. 2. From the rules described earlier, there will be two edge states flowing downwards between the metal boundary and the $C_{gap}=2$ crystal as indicated with arrows in Fig. 3d. These edge states will ``split'' at the junction with one edge state flowing away along the boundary between the $C_{gap}=1$ and the $C_{gap}=2$ crystal, and the other continuing along the metal and $C_{gap}=1$ crystal interface. 

\begin{figure}
\centering
		\includegraphics[width=8.5cm]{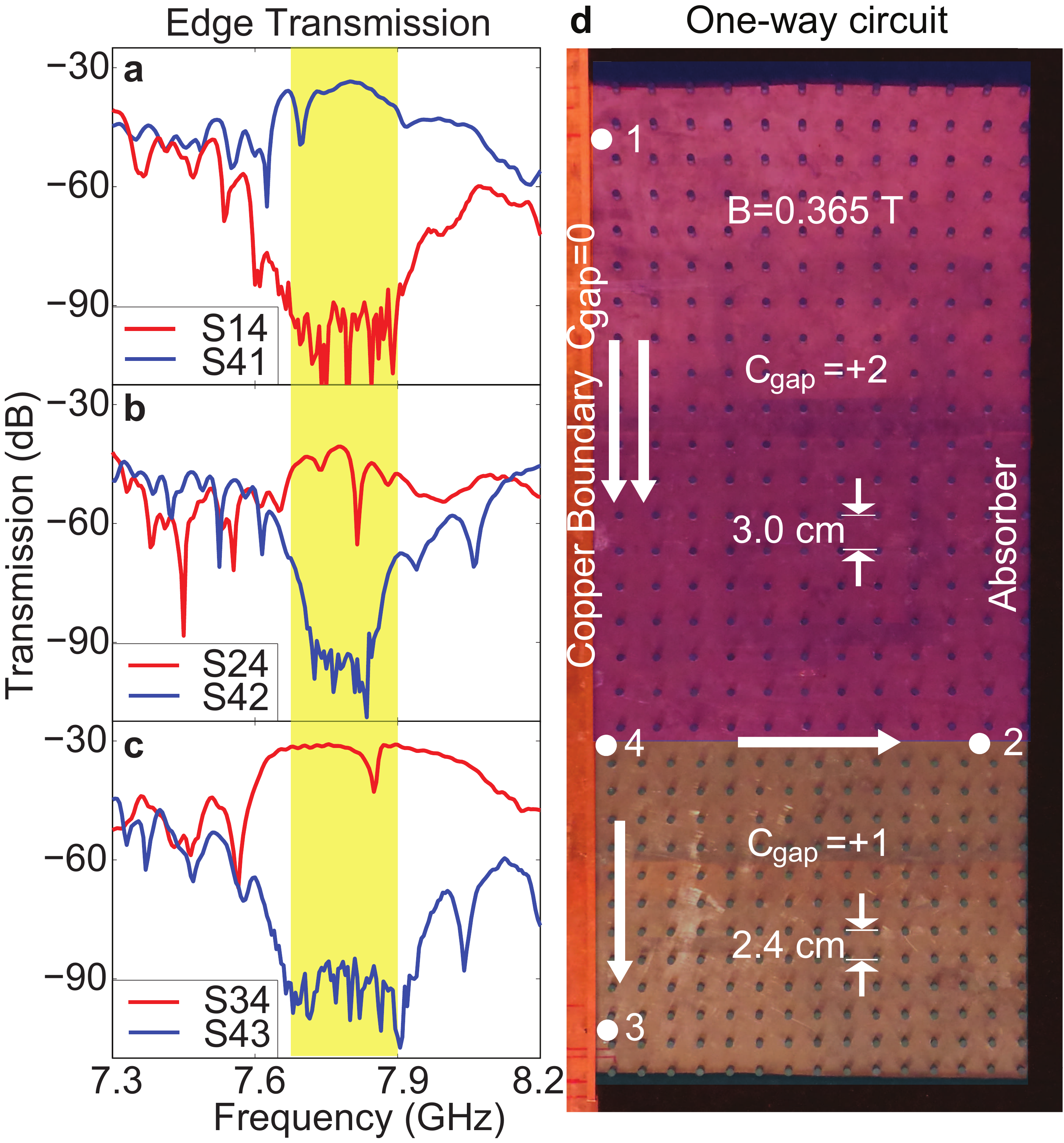} 
	\caption{Topological one-way circuit implemented using $C_{gap}=1$ (a=2.4 cm) and $C_{gap}=2$ (a=3.0 cm) photonic crystal. a)-c) Transmission plots showing edge transmission between antennas at 1,2, and 3 and antenna 4 located at the center. Shared bulk bandgap for $C_{gap}=1$ and $C_{gap}=2$ crystals is highlighted in yellow. d) Experimental configuration illustrating crystals with copper boundary ($C_{gap}=0$) on left and antenna locations 1-4. Arrows indicate the theoretical direction and the number of the edge states at each interface. The transmission data is consistent with predicted edge state directions, which confirms that the upper crystal has $C_{gap}>1$. }
	\label{fig:two-edge-modes}
\end{figure}

In Fig. 3a-c we present the transmission between ports 1-3 and a 4\textsuperscript{th} port located at the junction as labeled in Fig. 3d. The highlighted yellow region indicates the shared bandgap between the $C_{gap}=1$ crystal and the $C_{gap}=2$ crystal. For each of the measurements, it is clear that in one direction we have a strong bandgap, with a signal level at the noise floor of about -100 dB, while in the opposite direction there is 50 to 60 dB more transmission. These edge state directions are consistent with the theoretical predictions and prove the existence of $C_{gap}$>1 for the upper crystal. The results from Fig. 3 were obtained under an applied magnetic field of 0.365 T, although there was a window extending approximately from 0.32 T to 0.4 T where the $C_{gap}=2$ and $C_{gap}=1$ bandgaps from each crystal were well aligned (Supplementary Material). 


In conclusion, we experimentally constructed a square lattice ferrimagnetic photonic crystal with a bandstructure comprised of high $C_{gap}$ (-4 to 3) bandgaps and measured the dispersion relations of the multimode one-way edge waveguides. Fundamentally, having bandgaps with higher gap Chern numbers greatly expands the phases available for topological photonics. These results can potentially enable multi-mode one-way waveguides with high capacity and coupling efficiencies, as well as many other devices~\cite{he2010tunable,yang2013experimental,poo2011experimental,fu2011unidirectional}. A topological photonic circuit can also be made by interfacing photonic crystals of various $C_{gap}$, with one-way edge states combining together or splitting off at the junctions. Given the rapidly expanding literature on the QAHE and its analogs for $|C_{gap}|=1$~\cite{lu2014topological,hafezi2011robust,fang2012realizing,khanikaev2012photonic,kraus2012topological,rechtsman2013photonic,chen2014experimental}, many more avenues of research are now possible because of the greater range of topological phases that can be investigated. Our approach can be readily extended to other systems of Bosonic particles such as magnons~\cite{mook2014edge}, excitons~\cite{yuen2014topologically}, and phonons~\cite{yang2014topological,pai2015acoustics}.



\begin{acknowledgements}

We acknowledge Carl E Patton, Liang Fu, Bo Zhen, Ido Kaminer, Yichen Shen, Zheng Weng, and Hongsheng Chen for discussions, Ulrich Becker and Peter Fisher for assistance with the MIT cyclotron magnet, and Xiangdong Liang for numerical assistance. 
S. S. was supported by MIT Tom Frank Fellowship and NSF Fellowship.
L. L. was supported in part by the MRSEC Program of the NSF under Award No. DMR-1419807. 
Q. Y. was supported by the National Natural Science Foundation of China under Grants No. 61322501. 
Fabrication part of the effort was paid by the MIT S3TEC EFRC of DOE under Grant No. DE-SC0001299. Also supported in part by ARO through ISN, W911NF-13-D-0001. 

\end{acknowledgements}

\bibliographystyle{abbrv}

\clearpage

\end{document}